\documentclass{article}

\usepackage[final]{neurips_2020}

\usepackage[utf8]{inputenc} 
\usepackage[T1]{fontenc}    
\usepackage{hyperref}       
\usepackage{url}            
\usepackage{booktabs}       
\usepackage{amsfonts}       
\usepackage{amsmath}
\usepackage{nicefrac}       
\usepackage{microtype}      
\usepackage{graphicx}

\newcommand{\alphab}{\boldsymbol\alpha}
\newcommand{\betab}{\boldsymbol\beta}

\title{Using StyleGAN for Visual Interpretability of Deep Learning Models on Medical Images}

\author{%
  Kathryn Schutte\thanks{Corresponding author: \texttt{kathryn.schutte@owkin.com}}, \, Olivier Moindrot, \, Paul Hérent, \, Jean-Baptiste Schiratti, \, Simon Jégou \\
   \\Owkin, Inc. \\ New York, NY, USA\\
}

\begin{document}

\maketitle

\begin{abstract}
As AI-based medical devices are becoming more common in imaging fields like radiology and histology, interpretability of the underlying predictive models is crucial to expand their use in clinical practice. Existing heatmap-based interpretability methods such as GradCAM only highlight the location of predictive features but do not explain how they contribute to the prediction. In this paper, we propose a new interpretability method that can be used to understand the predictions of any black-box model on images, by showing how the input image would be modified in order to produce different predictions. A StyleGAN is trained on medical images to provide a mapping between latent vectors and images. Our method identifies the optimal direction in the latent space to create a change in the model prediction. By shifting the latent representation of an input image along this direction, we can produce a series of new synthetic images with changed predictions. We validate our approach on histology and radiology images, and demonstrate its ability to provide meaningful explanations that are more informative than GradCAM heatmaps. Our method reveals the patterns learned by the model, which allows clinicians to build trust in the model’s predictions, discover new biomarkers and eventually reveal potential biases.
\end{abstract}


\section{Introduction}
As of September 2020, the FDA had approved 64 AI-based medical devices~\citep{benjamens2020state}, and for the first time the Centers for Medicare \& Medicaid Services (CMS) approved the reimbursement of deep-learning powered stroke detector for brain CT scans~\citep{vizai}. 
The advances of deep learning in computer vision~\citep{krizhevsky2012imagenet} are especially promising in medical imaging fields such as radiology~\citep{ardila2019end}, histology~\citep{coudray2018classification}, dermatology~\citep{esteva2017dermatologist} or ophthalmology~\citep{gulshan2016development}.

While many deep learning techniques may provide state-of-the-art predictive performance, \emph{interpretable} deep learning models are necessary for regulatory approval, as their ability to explain their predictions can reveal potential biases and failure modes, as seen in the case of~\citep{ChestXray14}. Additionally, interpretable models also provide new opportunities for biomedical investigation, as evidenced in~\citep{courtiol2019deep}. Finally, such models are able to make inroads with medical experts, as their explainability helps build confidence in their utility~\citep{holzinger2019causability}.
As illustrated by the COVID-19 crisis~\citep{doi:10.1148/radiol.2020201491,doi:10.1148/radiol.2020200905,wang2020contrastive}, the go-to method for model interpretation in the medical imaging field is GradCAM~\citep{8237336}, which produces a coarse heatmap based on gradient intensity to identify which areas of the input image are responsible for the prediction.
However, these heatmaps only highlight the location of predictive features but do not explain how they contribute to the prediction.
In an image where information is diffuse, the heatmap cannot highlight any specific region so GradCAM is not sufficient to interpret the model predictions.

In this paper, we propose a new interpretability method that generates small synthetic transformations of the original image that would lead to different model predictions.
%
%
%
%
We train a generative model called StyleGAN~\citep{karras2019style,karras2020analyzing} and find the minimal modification in the latent space that changes the model prediction, which ensures that generated images remain as close as possible to the original image.
%
\citet{doi:10.1148/radiol.2018180887} explore a similar idea by using an older GAN algorithm to create heatmaps highlighting features of congestive heart failure, but their method cannot be applied to any black-box model.
\citet{fetty2020latent} manipulate three attributes of the StyleGAN latent space in order to enlarge datasets with synthetic images.
%
We validate our interpretability method on two different imaging modalities and demonstrate its ability to provide meaningful explanations of the predictions, and its potential to discover new biomarkers.


\section{Method}

We propose to create StyleGAN-generated visualizations that explain the predictions of a deep neural network in an interpretable manner.
Let $f$ be a classifier (e.g. a fully convolutional neural network) trained on a dataset $\mathcal{D} = \left( \mathbf{x}_i, y_i \right) \in \mathcal{X} \times \mathcal{Y}$, where $\mathcal{X}$ denotes a set of 2D images and $\mathcal{Y}$ a finite set of labels. Our method consists of three steps.
First, the images in $\mathcal{X}$ are used to train a StyleGAN2~\citep{karras2019style}, which is an improved GAN whose \emph{generator} $G \, : \, \mathcal{W} \, \rightarrow \, \mathcal{X}$ has a linearly disentangled intermediate \emph{latent space} $\mathcal{W} \subset \mathbb{R}^{512}$.
The generator $G$ is used to generate a set of synthetic images $\left( G\left( \mathbf{w}_i \right) \right)$, where the $\mathbf{w}_i$ are sampled in the latent space $\mathcal{W}$. 
Then, we train (using a Mean Squared Error loss) a ResNet50~\citep{he2016deep} \emph{encoder} $E \, : \, \mathcal{X} \, \rightarrow \, \mathcal{W}$ on the synthetic dataset $\left( \mathbf{w}_i, G\left( \mathbf{w}_i \right) \right)$ to retrieve the latent representation $\mathbf{w}_i$ from a generated image $G\left( \mathbf{w}_i \right)$.
Finally, a logistic regression classifier $\tilde{f}\left( \mathbf{w_i} \right) = \sigma\left( \alphab^{\top} \mathbf{w_i} + \betab \right)$ is trained on the latent space $\mathcal{W}$ to predict the estimated labels $\tilde{y}_i=f(G(\mathbf{w_i}))$ associated to each latent vector $\mathbf{w}_i \in \mathcal{W}$.
Given a new input image $\mathbf{x} \in \mathcal{X}$, our method translates the latent vector $\mathbf{w}=E\left(x\right)$ along the direction $\alphab$.
We can then create new images from the latent representation via $G\left( \mathbf{w} + \lambda \alphab \right)$ associated with a lower or a higher prediction depending on the value of $\lambda \in \mathbb{R}$. 


\begin{figure}
  \centering
  \includegraphics[width=\linewidth]{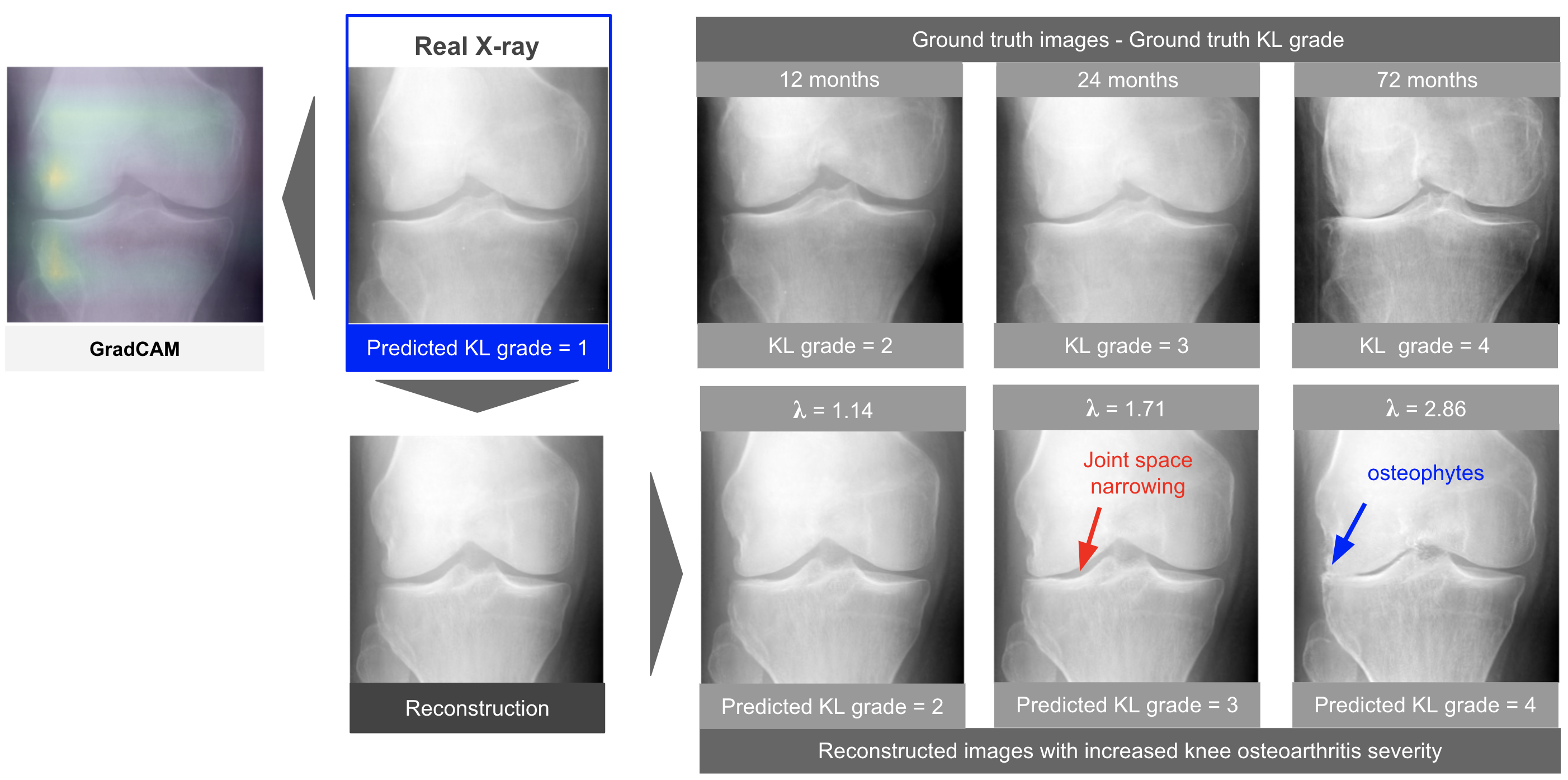}
  \caption{Our method applied to knee osteoarthritis severity prediction on an X-ray image. The input image is gradually modified to increase the osteoarthritis severity. The GradCAM heatmap is computed on the input image to compare both interpretability methods. X-rays of the patient's later visits are displayed to visually assess the clinical relevance of our method.}
  \label{radio}
\end{figure}

\section{Experiments}


\subsection{Knee osteoarthritis severity prediction on X-ray images}

We first demonstrate our method by explaining the predictions of an osteoarthritis severity predictor on X-ray images.
The dataset on which the predictor has been trained consists of 20,123 X-rays of patients suffering from knee osteoarthritis collected by the Osteoarthritis Initiative (OAI)~\citep{nevitt2006osteoarthritis}.
Each patient has one to eight 12-month follow-up X-rays, as well as associated clinical data, including the Kellgren and Lawrence (KL) grade~\citep{kohn2016classifications}.
The KL grade describes a degree of osteoarthritis severity and ranges from 0 to 4: grades 0 and 1 mean no or doubtful osteoarthritis, while grades 2 to 4 mean mild to severe osteoarthritis.

The image classifier $f$ is a ResNet50 trained on the multi-class prediction task.
To fit this multi-class setting to our method, we transform it to a binary classification task by pooling grades 0 and 1 versus grades 2 to 4.
The predictor $f$ obtains 89\% test AUC on this binary task, while $\tilde{f}$ obtains 80\% test AUC on the latent space.
Three radiologists evaluated the quality of the StyleGAN generator with a Turing test. They reach 58\% accuracy on average, showing that synthetic and real X-rays are almost indistinguishable.

In Figure~\ref{radio}, our interpretability method is applied to a real X-ray image. 
The GradCAM heatmap provides topographical information by showing that the osteoarthritis features are located in lateral femorotibial space.
Our method provides more than topographical information by showing the gradual emergence of the different osteoarthritis features as the KL grade increases, such as the joint space narrowing (red arrow) and osteophytes (blue arrow).
By comparing the synthetic evolution of the image to the real evolution of the patient at 12, 24 and 72 months after baseline, we observe that the direction found in the latent space corresponds to a biologically plausible osteoarthritis progression.

\subsection{Tumor detection on histology images of metastatic lymph nodes}

We apply the same method to histology images, to explain the predictions of a metastasis detector on Camelyon16~\citep{bejnordi2017diagnostic}.
The dataset contains 224,166 patch images from breast cancer lymph node whole-slide images, each with a binary label indicating the presence of tumor cells.

\begin{figure}
  \centering
  \includegraphics[width=\linewidth]{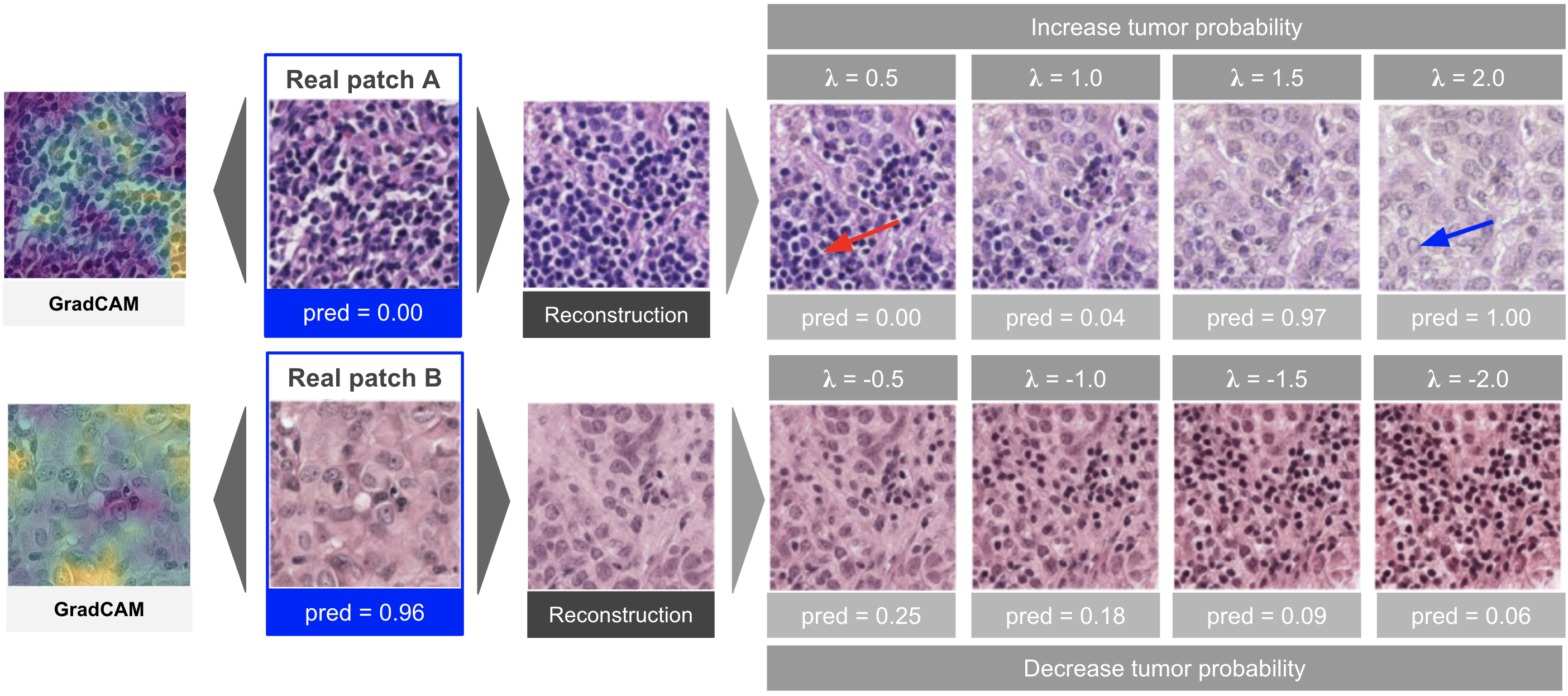}
  \caption{Our method applied to tumor probability prediction on two histology tiles of metastatic lymph nodes. The input image is gradually modified to increase (on patch A) or decrease (on patch B) the tumor probability. The GradCAM heatmap is computed on the input images to compare both interpretability methods.}
  \label{fig:histo}
\end{figure}

The image classifier $f$ is a ResNet50 trained on this dataset, obtaining 92\% test AUC, while the latent predictor $\tilde{f}$ reaches 95\% test AUC.
Figure~\ref{fig:histo} shows our interpretability method on two images: patch B contains tumoral cells while patch A does not.
The GradCAM heatmaps are not relevant here because the informative features are spread over the entire image.
On the contrary, our approach reveals clinically relevant features.
On patch A, it shows the appearance of tumor cells (blue arrow) and the disappearance of lymphocytes (red arrow) as the tumor probability increases, and inversely on patch B.

We can see that the encoder-decoder model is not able to perfectly reconstruct histology images,  as opposed to knee X-rays. A possible explanation is that the StyleGAN model does not generate images that are under-represented in the training set. This issue is highlighted in this particular use-case as there is more variability in the histology images than in the  knee X-ray images. Recently,~\citet{yu2020inclusive} propose to overcome this data coverage challenge by harmonizing adversarial training with reconstructive generation.

\section{Conclusion}

In this study we explored the potential of StyleGANs to explain the predictions of black-box models on medical images.
%
Although heatmap-based methods dominate the interpretability field, they only highlight the localization of predictive features in the image.
Our method provides an intuitive way for medical researchers to understand \emph{where} are located the predictive features in the image and \emph{how} they impact the prediction by showing modified views of the input image that would produce different predictions. This method shows how the model learned to solve the prediction task which allows clinicians to build trust in the model's predictions, discover new biomarkers and eventually reveal potential biases. In both experiments, our method proved that the models learned clinically relevant features.

\subsubsection*{Acknowledgments}

We thank Eric W. Tramel for his valuable feedback on the manuscript. We thank the three radiologists Eric Pessis, François Legoux and Thibaut Emorine for their participation in the Turing Test. 

\newpage

\bibliographystyle{apalike}
\bibliography{biblio}

\end{document}